\documentclass[prb,preprint,groupedaddress, showkeys,showpacs]{revtex4}

\usepackage{graphicx}
\usepackage{fancyhdr}
\begin{document}


\title{Phase shift induces currents in a periodic tube}


\author{Bao-quan  Ai$^{a}$}\email[Email: ]{aibq@scnu.edu.cn}
\author{Liang-gang Liu$^{b}$}

\affiliation{$^{a}$ Institute for Condensed Matter Physics, School
of Physics and Telecommunication Engineering, South China
Normal University, 510006 GuangZhou, China.\\
$^{b}$ The Faculty of Science  Technology , Macau University of
Science and Technology, Macao.}


\date{\today}
\begin{abstract}
\indent The average current of an overdamped Brownian particle
moving along the axis of a three-dimensional periodic tube is
investigated in the presence of a symmetric potential and a
temporally symmetric unbiased external force. Reduction of the
spatial dimensionality from two or three physical dimensions to an
effective one dimensional system entails the appearance of not
only an entropic barrier but also an effective diffusion
coefficient. We find that the phase shift between the tube shape
and the potential can break the symmetry of the effective
potential and can induce net currents. Under optimal condition,
the current as a function of temperature and the phase shift
possesses many extrema of alternating signs. The current may
reverse its direction several times when temperature or the phase
shift is changed. Our model is to describe the movement of
molecular motors along microtubule.
\end{abstract}

\pacs{05. 60. Cd, 05. 40. Jc, 87. 10. +e }
\keywords{Phase shift, molecular motors, Current reversal}



\maketitle

\indent

\section {Introduction}
\indent Molecular motors are protein molecules that can convert
chemical energy, usually in the form of adenosine triphosphate
(ATP), into mechanical forces and motion. Most organisms have many
different motors that are specialized for particular purposes such
as cell division, cell crawling, cell shape maintenance and
movements of internal organelles \cite{1,2}. Three different
linear families of molecular motors have been identified \cite{3}:
Kinesins and dyneins that move along microtubule, myosins that
move along actin filaments. From a theoretical point of view,
molecular motors are microscopic objects that move
unidirectionally along periodic structures. The problem of this
unidirectionality belongs to a larger class of such problems
involving rectifying processes at small scale. Noise-induced
transport phenomena play a crucial role in molecular
motors\cite{4,5,6,7,8}. A molecular motor is usually described by
a ratchet system which is generally defined as a system that is
able to transport particles in a periodic structure with nonzero
macroscopic velocity in the
absence of macroscopic force on average.\\
\indent In these systems, directed Brownian motion of particles is
generated by nonequilibrium noises in the absence of any net
macroscopic forces and potential gradients. Typical examples are
rocking ratchets \cite{7,8}, flashing ratchets \cite{9}, diffusion
ratchets \cite{10}, correlation ratchets \cite{7,8} and
white-shot-noise ratchets \cite{5}. In all these studies, the
potential is asymmetric in space. It has shown that a
unidirectional current can also appear for spatially symmetric
potentials if the external random force is either asymmetric or
spatially dependent. If spatially periodic structures are exposed
to an additive Poissonian white shot noise, a macroscopic current
occurs even in the absence of spatial
asymmetry \cite{5}.\\
\indent Most studies have revolved around the energy barrier. The
nature of the barrier depends on which thermodynamic potential
(internal energy or Helmholtz free energy) varies when passing
from one well to the other, and its presence plays an important
role in the dynamics of the system. Whereas energy barriers are
more frequent in problem of solid-state physics (metals and
semiconductors, coupled Josephson junction and photon crystal),
entropy barriers are often encountered in soft-condensed matter
and biological systems. Entropy barriers may appear when
coarsening the description of a complex system in order to
simplify its dynamics. Reguera and co-workers \cite{12} used the
mesoscopic nonequilibrium thermodynamics theory to derive the
general kinetic equation of the motor system and analyzed in
detail the case of diffusion in a domain of irregular geometry in
which the presence of the boundaries induces an entropy barrier
when approaching the dynamics by a coarsening of the description.
In their recent work \cite{13}, they studied the current and the
diffusion of a Brownian particle moving in a symmetric channel
with a biased external force. They found that temperature dictates
the strength of the entropic potential, and thus an increasing of
temperature leads to a reduction of the current.

\indent The focus of previous works on current were limited to the
case of energy barriers in the presence of asymmetry in space or
time. The present work is extended  to the study to case of
entropic barriers in the presence symmetry in space and time. We
emphasize on finding how phase shift between the tube shape and
the potential can induce a net current. Our model is to describe
the movement of kinesins and dyneins along  microtubule.

\section {The current induced by a phase shift}

\indent Phenomena of current reversals may be of interest in
molecular motors\cite{14}. kinesins and dyneins move along tubulin
filaments towards their plus and minus extremities, respectively.
It is well known that the two current reversals effect allows one
pair of motor proteins (two members)to move simultaneously in
opposite directions along the microtubule inside the eukaryotic
cells. One member of the pair moves one way along the tube while
the other member of the pair moves in the opposite direction. In
order to describe the movement of kinesins and dyneins along
tubulin, a periodic tube model is proposed. We consider a Brownian
particle moving in a symmetric periodic tube [Fig. 1] in the
presence of a temporally symmetric unbiased external force. Its
overdamped dynamics is described by the following Langevin
equations written in a dimensionless form \cite{12,13},
\begin{equation}\label{}
    \eta\frac{dx}{dt}=-\frac{\partial U(x)}{\partial x}+F(t)+\sqrt{\eta
    k_{B}T}\xi_{x}(t),
\end{equation}
\begin{equation}\label{}
    \eta\frac{dy}{dt}=\sqrt{\eta
    k_{B}T}\xi_{y}(t),
\end{equation}
\begin{equation}\label{}
    \eta\frac{dz}{dt}=\sqrt{\eta
    k_{B}T}\xi_{z}(t),
\end{equation}
where $x$, $y$, $z$  are the three-dimensional (3D) coordinates,
$\eta$ is the friction coefficient of the particle, $k_{B}$ is the
Boltzmann constant, $T$ is the absolute temperature and
$\xi_{x,y,z}(t)$ is the Gaussian white noise with zero mean and
correlation function:
$<\xi_{i}(t)\xi_{j}(t^{'})>=2\delta_{i,j}\delta(t-t^{'})$ for
$i,j=x, y, z$. $<...>$ denotes an ensemble average over the
distribution of noise. $\delta(t)$ is the Dirac delta function.
Imposing reflecting boundary conditions in the transverse
direction ensures the confinement of the dynamics within the tube,
while periodic boundary conditions are enforced along the
longitudinal direction for the reasons noted above. $F(t)$ is a
temporally symmetric unbiased external force along the $x$
direction which is imparted as the result of energy gained via ATP
hydrolysis and satisfies\cite{5,6,7,8}
\begin{equation}\label{}
 F(t)=\left\{
\begin{array}{ll}
   F_{0},& \hbox{$n\tau\leq
t<n\tau+\frac{1}{2}\tau$};\\
   -F_{0} ,&\hbox{$n\tau+\frac{1}{2}\tau<t\leq
      (n+1)\tau$},\\
\end{array}
\right.
\end{equation}
where $\tau$ is the period of the unbiased force and $F_{0}$ is
its magnitude.

\indent $U(x)$ is a symmetric potential with periodicity $L$ which
is from the track of molecular motors [Fig. 1],
\begin{equation}\label{}
    U(x)=Q\sin(\frac{2\pi x}{L}),
\end{equation}
where $Q$ is the amplitude of the potential. The shape of the tube
(describing the structure of molecular motors) is described by its
radius,
\begin{equation}\label{}
    \omega(x)=a\sin(\frac{2\pi x}{L}+\phi)+b,
\end{equation}
where $a$ is the parameter that controls the slope of the tube,
$\phi$ is the relative phase shift between $\omega(x)$ and $U(x)$.
The radius at the bottleneck is $b-a$.

\indent The movement equation of a Brownian particle moving along
the axis of the 3D (or 2D) tube can be described by the
Fick-Jacobs equation \cite{12,13,15,16} which is derived from the
3D (or 2D) Smoluchowski equation after elimination of $y$ and $z$
coordinates by assuming equilibrium in the orthogonal directions.
Reduction of the spatial dimensionality from two or three physical
dimensions to an effective one dimensional system  may involve not
only the appearance of an entropic barrier, but also the effective
diffusion coefficient. When $|\omega^{'}(x)|<<1$, the effective
diffusion coefficient is \cite{12,13}
\begin{equation}\label{}
    D(x)=\frac{D_{0}}{[1+\omega^{'}(x)^{2}]^{\alpha}},
\end{equation}
where $D_{0}=k_{B}T/\eta$ and $\alpha=1/3$ and $1/2$ for two and
three dimensions, respectively. The prime stands for the
derivative with respect to the space variable $x$.

\indent Consider the effective diffusion coefficient and the
entropic barrier, the dynamics of a Brownian particle moving along
the axis of the 3D (or 2D) tube can be described by
\cite{12,13,15,16}
\begin{equation}\label{}
    \frac{\partial P(x,t)}{\partial t}=\frac{\partial}{\partial x}[D(x)\frac{\partial P(x,t)}{\partial
    x}+\frac{D(x)}{k_{B}T}\frac{\partial A(x,t)}{\partial x}P(x,t)]=-\frac{\partial j(x,t)}{\partial
    x},
\end{equation}
where we define a free energy $A(x,t):=E-TS=U(x)-F(t)x-Tk_{B}\ln
h(x)$, here $E=U(x)-F(t)x$ is the energy, $S=k_{B}\ln h(x)$ is the
entropy, $h(x)$ is the dimensionless width $2\omega(x)/L$ in two
dimensions, and the dimensionless transverse cross section
$\pi[\omega(x)/L]^{2}$ of the tube in three dimensions. Here, we
define a effective potential along the $x$ coordinate,
\begin{equation}\label{}
    U_{eff}(x)=U(x)-Tk_{B}\ln h(x).
\end{equation}
\indent $j(x,t)$ is the probability current density. $P(x,t)$ is
the probability density for the particle at position $x$ and at
time $t$. It satisfies the normalization condition
$\int_{0}^{L}P(x,t)dx=1$ and the periodicity condition
$P(x,t)=P(x+L,t)$.

\indent If $F(t)$ changes very slowly with respect to $t$, namely,
its period is longer than any other time scale of the system,
 there exists a quasisteady state. In this case, by following the  method in  \cite{4,5,6,7,8,9,10,11,12,13}, we can obtain the
current
\begin{equation}\label{}
    j(F(t))=\frac{k_{B}T[1-\exp(-\frac{F(t)L}{k_{B}T})]}{\int_{0}^{L}h(x)\exp[\frac{-U(x)+F(t)x}{k_{B}T}]dx\int_{x}^{x+L}[1+\omega^{'}(y)^{2}]^{\alpha}h^{-1}(y)\exp[\frac{U(y)-F(t)y}{k_{B}T}]dy}.
\end{equation}

\indent For the given external force (Eq. (4)), the average
current \cite{7,8} is
\begin{equation}\label{9}
    J=\frac{1}{\tau}\int_{0}^{\tau}j(F(t))dt=\frac{1}{2}[j(F_{0})+j(-F_{0})].
\end{equation}

\section{Results and discussions}
\indent  Because the results  from two and three dimensions are
very similar, for the convenience of physical discussion, we now
mainly investigate the current in three dimensions with $k_{B}=1$,
$\eta=1$ and $L=2\pi$. The results are shown in Figs. 2-5.


\indent Figure 2 shows the current contours on $T-\phi$ plane.
There are two hills and two vales in the plot. The current is
negative in area A and C, positive in area B and D. In our model,
the potential and the tube shape in space and the external force
in time are symmetric. It is obvious that the phase shift between
the potential and the tube shape leads to a net current. However,
the current is always zero at $\phi=0$, $\phi$ and $2\pi$. When
$0<\phi<\phi_{a}$ and $\phi_{b}<\phi<2\pi$, the particle will
undergo a hill and a vale on increasing temperature, namely, the
current reverses its direction. When temperature is increased, the
particle will undergo only a hill or a vale for
$\phi_{a}<\phi<\pi$ and $\pi<\phi<\phi_{b}$. The current is always
zero at $\phi=\pi$. Because of the numerical trouble near
$\phi=1.0\pi$ and low temperature, the numerical results at
$\phi=1.0\pi$ and low temperature meet a little error. However,
from the effective potential (Eq. (9)), It is easy to find that
the effective potential is symmetric and the current is zero at
$\phi=1.0\pi$. So the line at $\phi=\pi$ should be vertical.


\indent The current $J$ as a function of temperature $T$ is shown
in Fig. 3 for different values of $\phi$. When $T\rightarrow 0$,
the particle cannot pass the barrier, so the current $J$ tends to
zero. When $T\rightarrow \infty$, the thermal noise is very large,
so the ratchet effect disappears and the current $J$ approaches
zero, also. There exist optimized temperatures at which the
current $J$ takes its maximal or minimal values. When
$\phi=0.2\pi$ ($0<\phi<\phi_{a}$) and
$1.8\pi$($\phi_{b}<\phi<2\pi$), current reversals may occur on
increasing temperature.  The current is always positive and
negative for $\phi=0.7\pi$ ($\phi_{a}<\phi<\pi$) and $1.4\pi$
($\pi<\phi<\phi_{b}$), respectively. It is easy to see from the
Figs. 2 and 3 that the current on increasing temperature may
reverse its direction for $0<\phi<\phi_{a}$ and
$\phi_{b}<\phi<2\pi$ and is always positive at all temperatures
for $\phi_{a}<\phi<\pi$ or negative for $\pi<\phi<\phi_{b}$.


\indent In order to illustrate the current (see Fig. 3) in detail,
the effective potential $U_{eff}(x)$ along the $x$ coordinate is
plotted in Fig. 4. From the previous works\cite{5,6,7,8,9,10,11},
we can obtain that there is no current when the effective
potential is symmetric. When the right of the potential in a
period is steeper than the left, the current is positive. The
current is negative when the left is steeper. In Fig 4
(a)($\phi=0.2\pi$), the left is steeper than the right (inducing a
positive current) at low temperatures. However, when temperature
is increased, the right may be steeper than the left (inducing a
negative current) at high temperatures. For the case $\phi=1.8\pi$
(Fig. 4(d)), the current is negative at low temperatures and
positive at high temperatures. Therefore, the current can change
its direction on increasing temperature for these two cases
($\phi=0.2\pi$ and $1.8\pi$). In Fig. 4 (b) ($\phi=0.7\pi$), the
right is always steeper than the left, so the current is always
positive.  The left is always steeper than the right at
$\phi=1.4\pi$ (Fig. 4 (c)), so the current is always negative. In
these two cases, no current reversal occurs.


\indent Figure 5 shows the current $J$ as a function of the phase
shift $\phi$. There is no net current at $\phi=0$, $\pi$ and
$2\pi$ for all temperatures. There are two peaks and two vales in
the curve on increasing the phase shift at $T=0.2$ (low
temperature). The current can change its direction even three
times when the phase shift is changed. At high temperature
($T=0.5$), only one peak and one vale appear and the current
reverses its direction at $\phi=\pi$.

\section{Concluding Remarks}
\indent In present work, we study the transport of a Brownian
particle moving along the axis of three-dimensional periodic tube
in the presence of a temporally symmetric unbiased force and a
symmetric potential. The movement equation can be described by the
Fick-Jacobs equation which is derived from 3D (or 2D) Smoluchowski
equation after elimination of $y$ and $z$ coordinates. Reduction
of the coordinates may induce the appearance of an entropic
barrier and the effective diffusion coefficient. The current is
achieved by solving one-dimensional Fick-Jacobs equation.\\
\indent It is found that the phase shift between the tube shape
 and the potential may induce a net current even when the potential,
 the tube shape and the external force are symmetric. The
 symmetry of the effective potential can be broken by changing
 temperature or the phase shift. When  $0<\phi<\phi_{a}$ and
 $\phi_{b}<\phi<2\pi$, current reversal may occur on
 increasing temperature. However,  the current is always positive
 for $\phi_{a}<\phi<\pi$ or negative for $\pi<\phi<\phi_{b}$. No current occurs at $\phi=0$, $\pi$ and
$2\pi$ for all temperatures. At low temperatures the current may
reverse its direction three times on increasing the phase shift.
At high temperatures the current changes its direction one time at
$\phi=\pi$.

\indent Our tube model is to describe molecular motors, especially
for kinesins and dyneins along microtubule. The potential
describes the track of the motor. The external force $F(t)$
depicts the stroke force due to ATP hydrolyzing. The shape of the
tube describes the structure of the motor. The phase shift between
the tube shape and the potential may induce a net current. Our
results can explain current reversals in molecular motors. Several
biological molecular motors, for instance kinesin and non-claret
disjunctions, belonging to the same superfamily of motor proteins
move towards opposite ends of the microtubule \cite{7}. This can
be explained by current reversals

\indent The results we have presented have broad applications in
many processes \cite{12}, such as molecular motors
  movement through the microtubule in the absence of any net macroscopic
  forces \cite{17},  ion transport through ion channels \cite{18}, motion of polymers
  subjected to rigid constraints \cite{19}, drug release \cite{20}and polymer
  crystallization \cite{21}. In these systems, a directed-transport with entropic or energy barriers can be
  obtained in the absence of any net macroscopic forces or in the presence of the unbiased forces.

\section{ACKNOWLEDGMENTS}
The authors thank Prof. Masahiro NAKANO for helpful discussions.
The work is supported by the National Natural Science Foundation
of China under Grant No. 30600122 and  GuangDong Provincial
Natural Science Foundation under Grant No. 06025073.

\newpage
\section {Caption list}
\baselineskip 0.4 in

FIG. 1. The upper is schematic diagram of tube with periodicity
$L$. The shape is described by the radius of the tube
$\omega(x)=a\sin(\frac{2\pi x}{L}+\phi)+b$. The bottom is the
potential along the $x$ coordinate with periodicity $L$,
$U(x)=Q\sin(\frac{2\pi x}{L})$.\\

FIG. 2. Current contours on $T-\phi (\pi)$ plane at $Q=0.5$,
$a=\frac{1}{2\pi}$, $b=\frac{1.5}{2\pi}$, $\alpha=1/2$ and
$F=0.5$. The current is negative in area A and C, positive in area
B and D. The solid line denotes zero current. $\phi_{a}$ (0.3817)
and $\phi_{b}$ (1.6035) are the intersectant point between the
$\phi$-axis and the zero-current line near area A and D,
respectively.\\

FIG. 3. Current $J$ vs temperature $T$ for different values of
$\phi$ at $Q=0.5$, $a=\frac{1}{2\pi}$, $b=\frac{1.5}{2\pi}$,
$\alpha=1/2$ and $F_{0}=0.5$.\\

FIG. 4. The effective potential $U_{eff}(x)$ along the $x$
coordinate for different values of $T$. From the bottom to the
upper the temperature is increased. (a) $\phi=0.2\pi$;
(b)$\phi=0.7\pi$; (c)$\phi=1.4\pi$; (d)$\phi=1.8\pi$.\\

FIG. 5. Current $J$ vs the phase shift $\phi$ for different
temperature at $Q=0.5$, $a=\frac{1}{2\pi}$, $b=\frac{1.5}{2\pi}$,
$\alpha=1/2$ and $F_{0}=0.5$.

\newpage
\begin{figure}[htbp]
  \begin{center}\includegraphics[width=10cm,height=6cm]{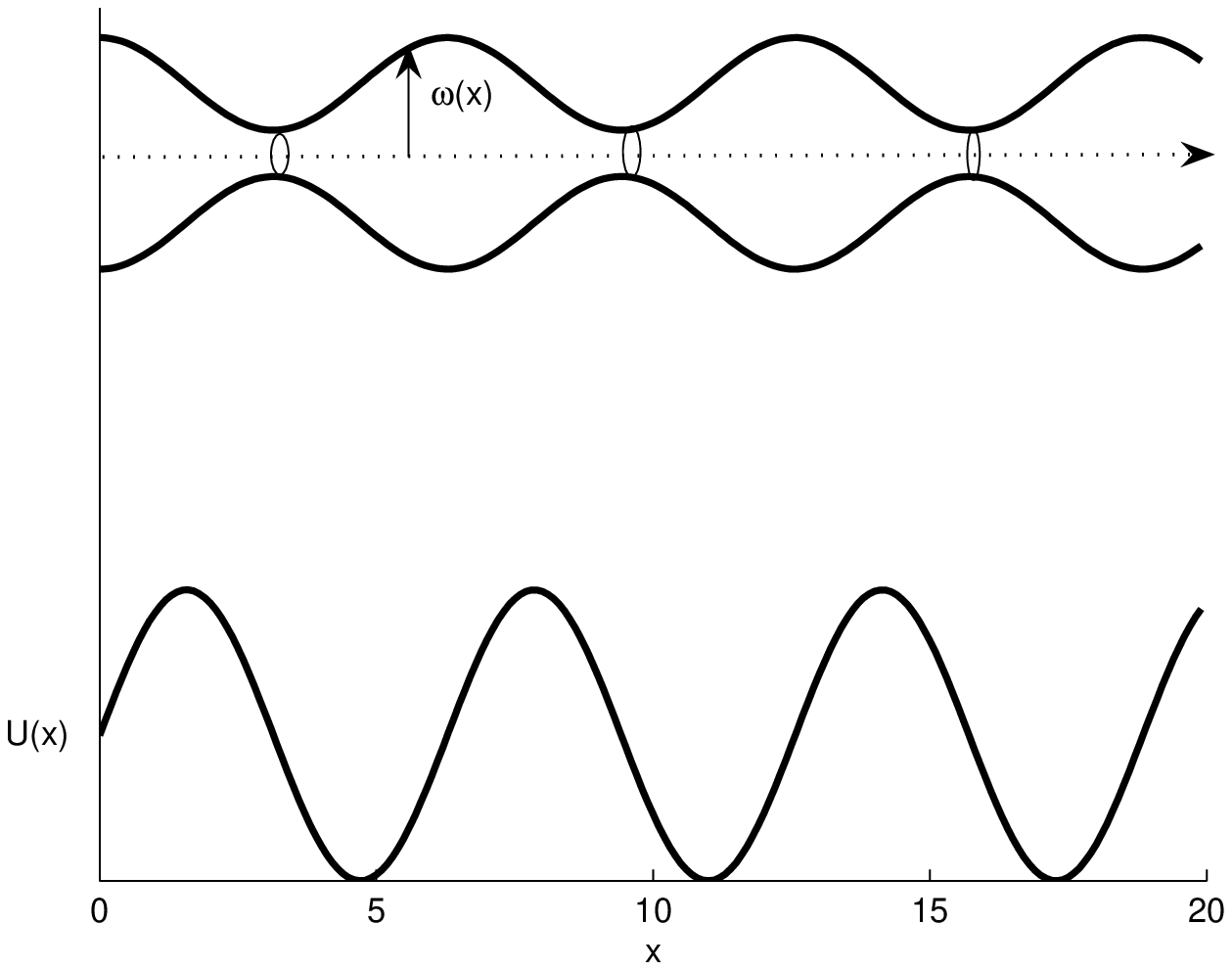}
  \caption{}\label{1}
\end{center}
\end{figure}
\newpage
\begin{figure}[htbp]
  \begin{center}\includegraphics[width=14cm,height=12cm]{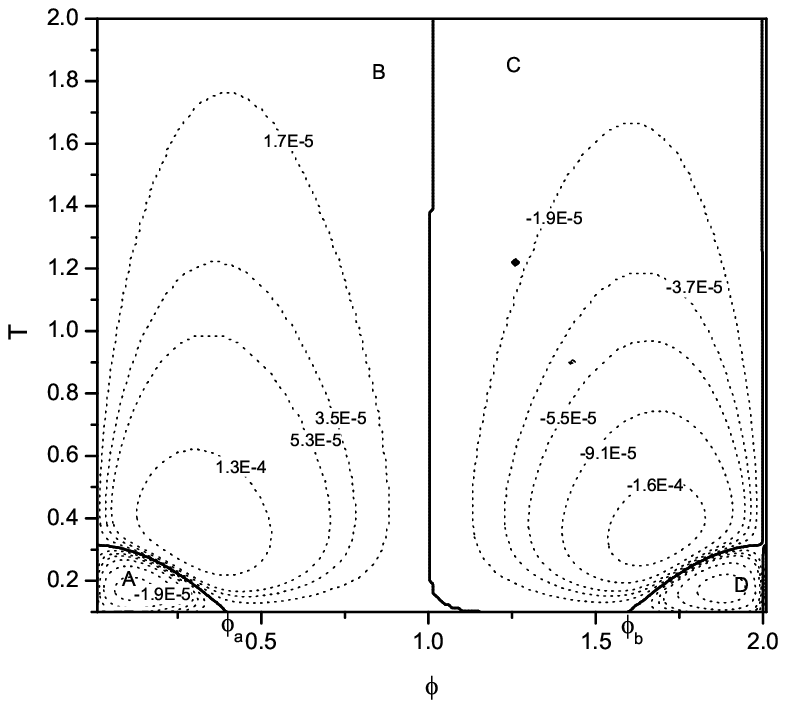}
  \caption{  }\label{1}
\end{center}
\end{figure}
\newpage
\begin{figure}[htbp]
  \begin{center}\includegraphics[width=10cm,height=8cm]{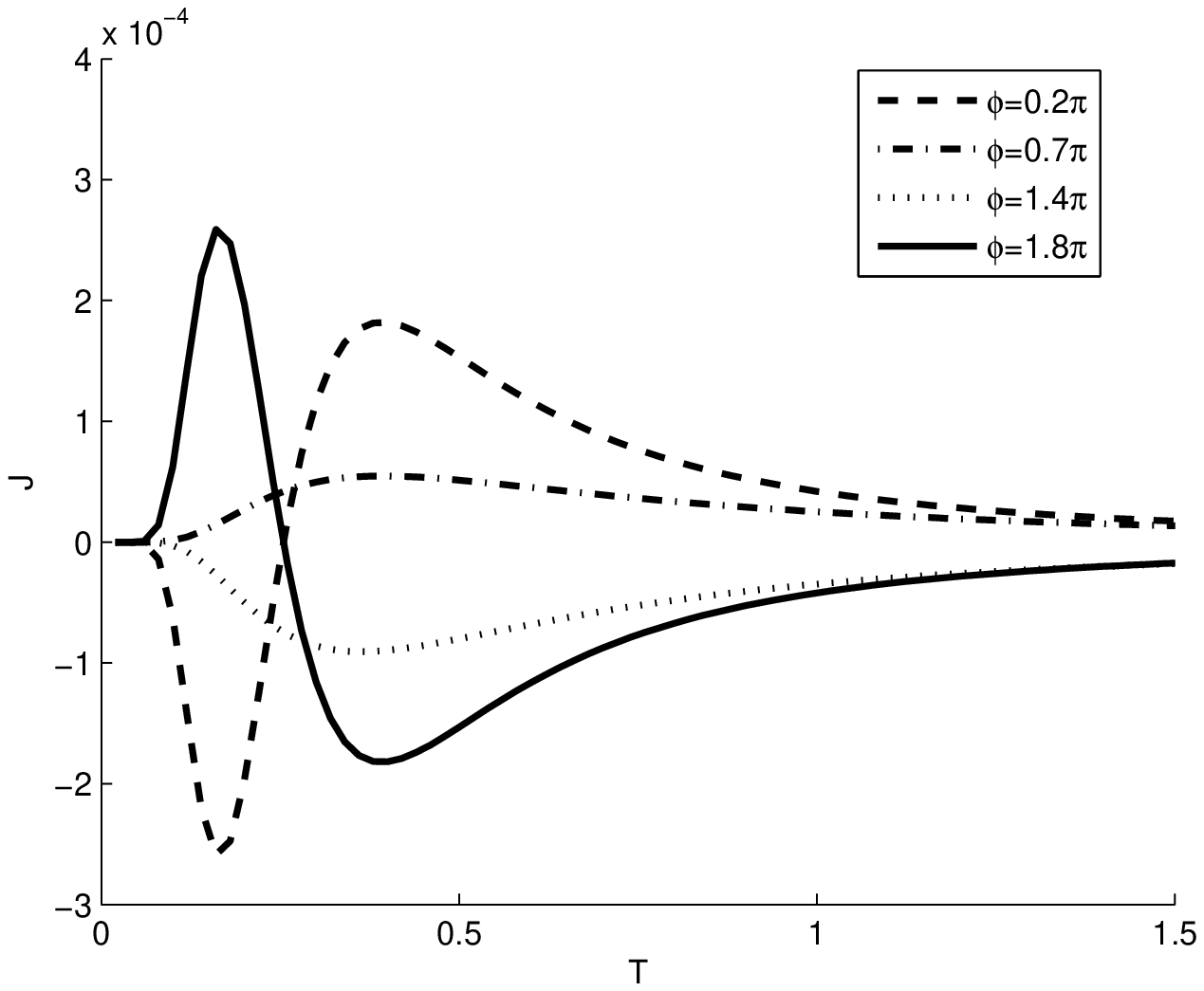}
  \caption{}\label{1}
\end{center}
\end{figure}
\newpage
\begin{figure}[htbp]
  \begin{center}\includegraphics[width=6cm,height=4cm]{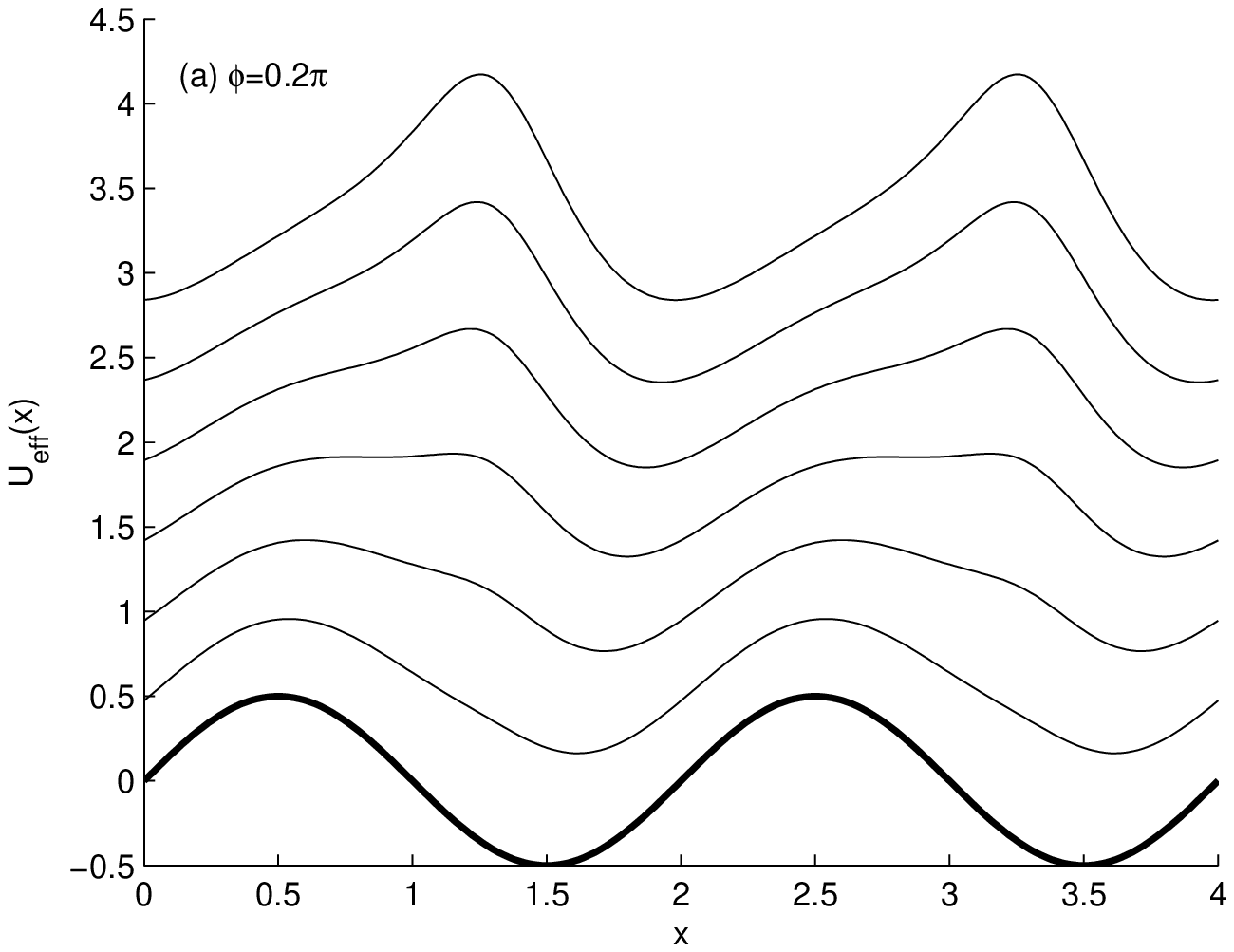}
  \includegraphics[width=6cm,height=4cm]{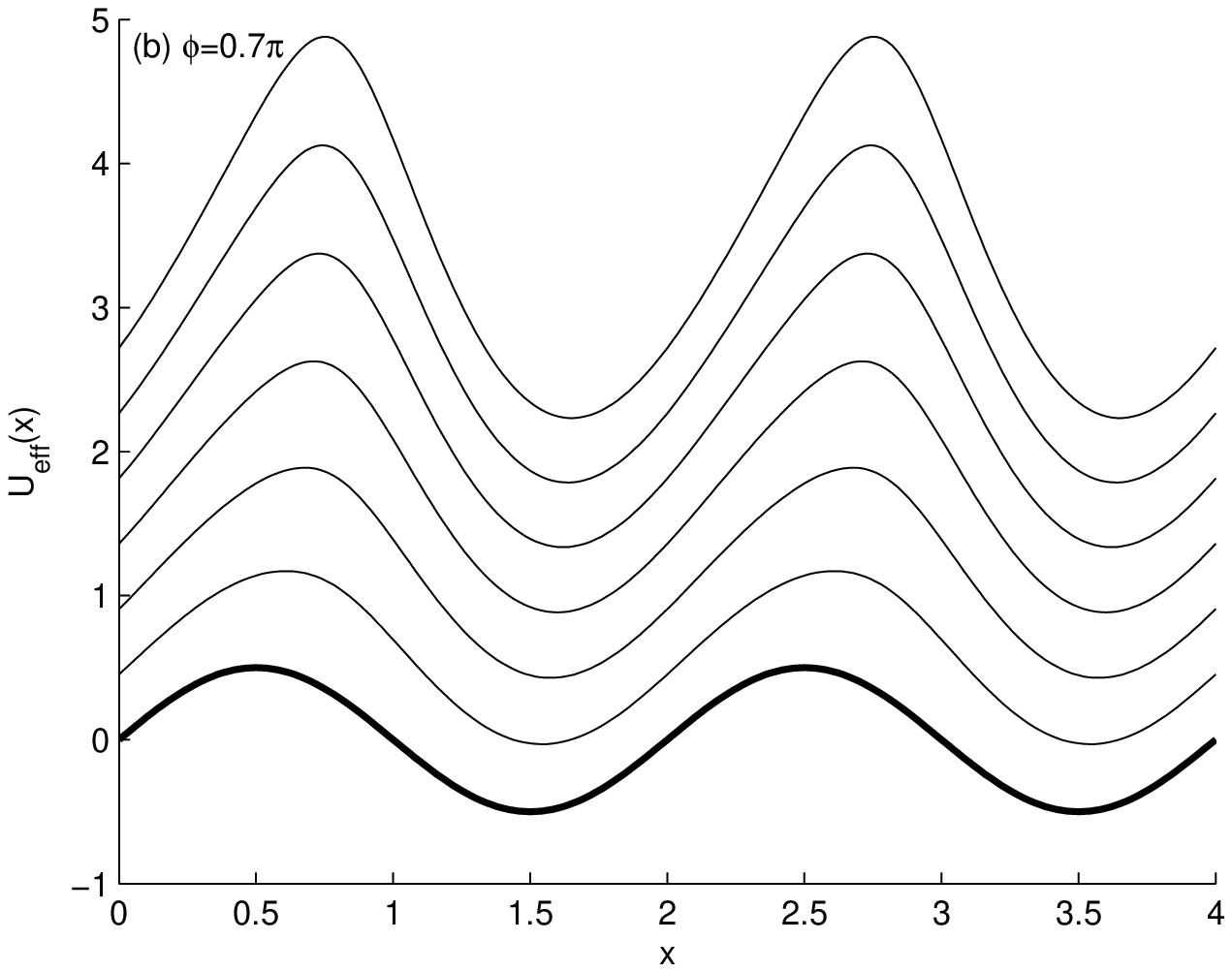}
  \includegraphics[width=6cm,height=4cm]{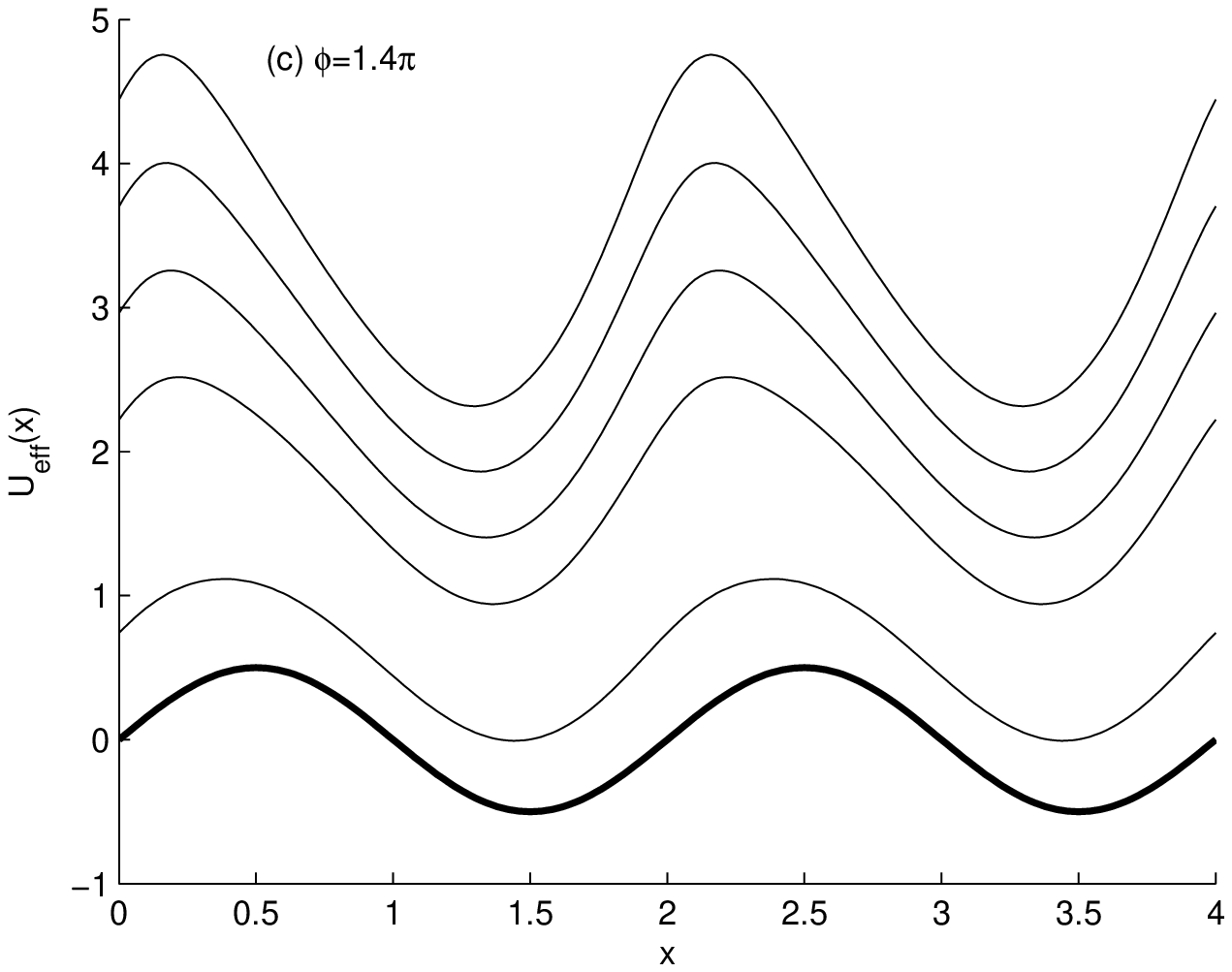}
  \includegraphics[width=6cm,height=4cm]{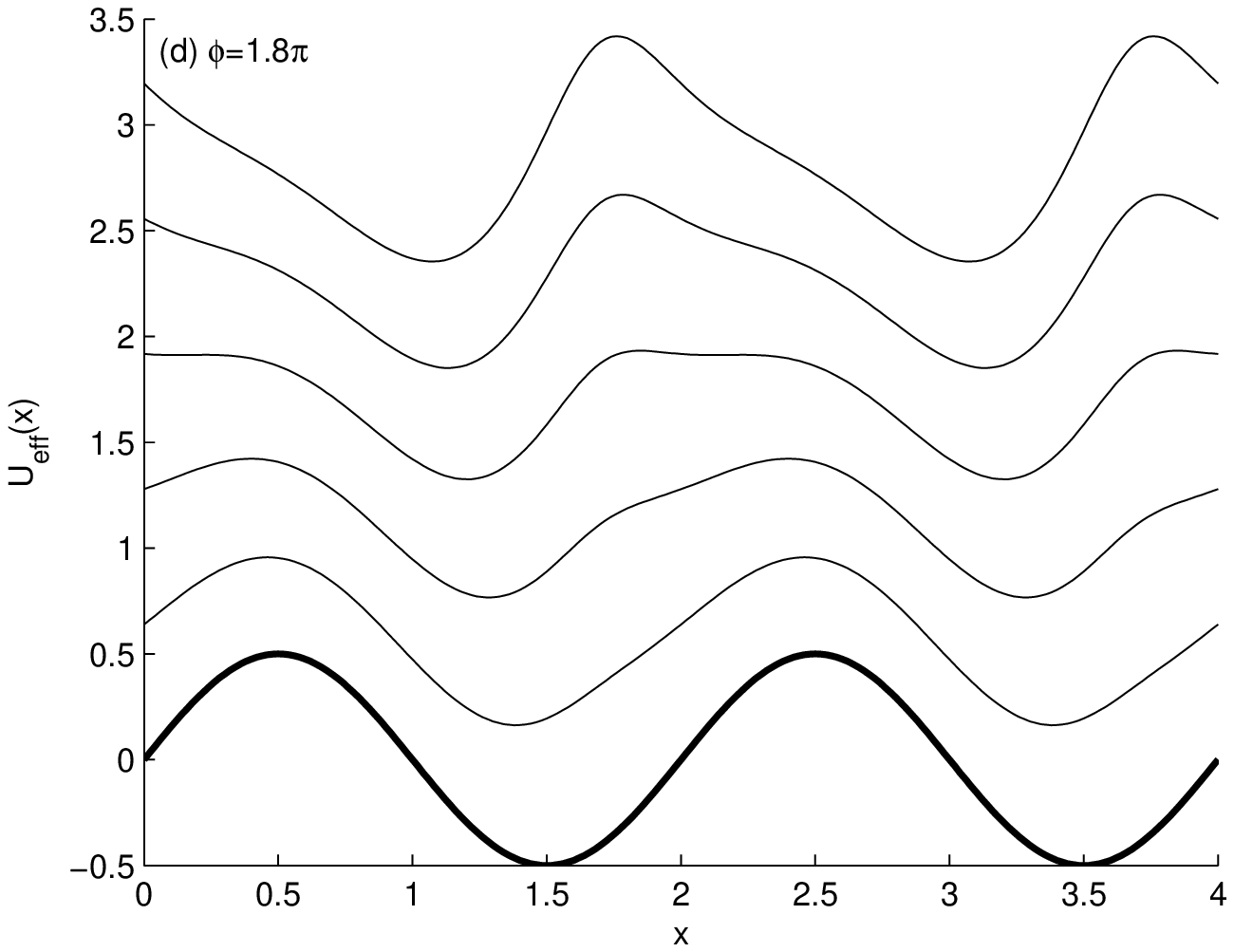}
  \caption{}\label{1}
\end{center}
\end{figure}
\newpage
\begin{figure}[htbp]
 \begin{center}\includegraphics[width=10cm,height=8cm]{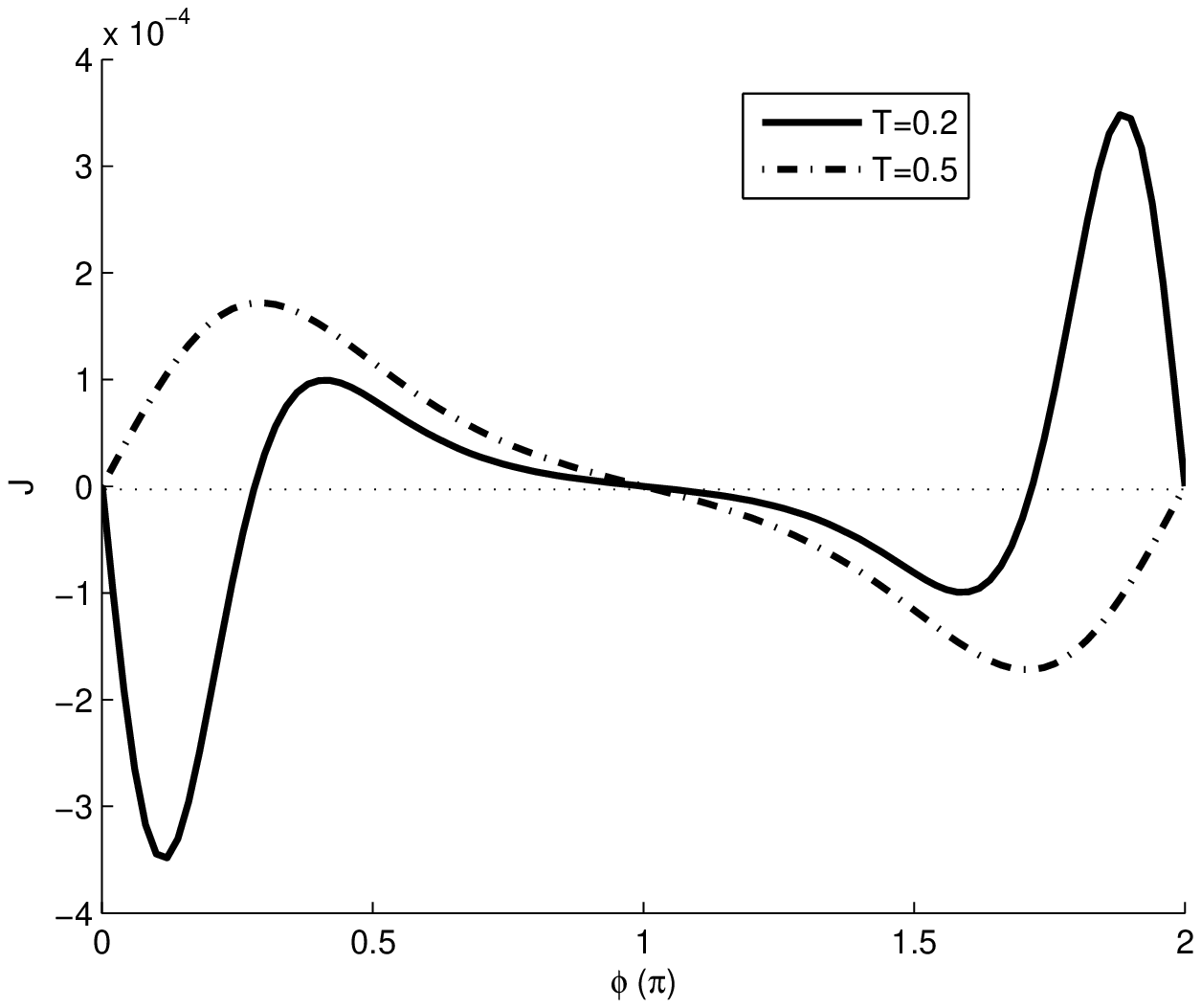}
  \caption{}\label{1}
\end{center}
\end{figure}
\newpage


\begin{thebibliography}{}
\bibitem{1}D. Keller and C. Bustamante, Biophysical Journal  78,
541 (2000).
\bibitem{2}C. Bustamante, D. Keller and G. Oster, Acc. Chem. Res.
34, 412 (2001).
\bibitem{3}F. Julicher, A. Ajdari and J. Prost, Rev. Mod. Phys.
69, 1269 (1997).
\bibitem{4}L. P. Faucheux et al., Phys. Rev. Lett. 74 (1995) 1504.
\bibitem{5}J. Luczka, R. Bartussek and P. Hanggi, Europhysics
Letters 31 (8), 431-436 (1995).
\bibitem{6}J. L. Mateos, Phys. Rev. Lett. 84, 258 (2000).
\bibitem{7}B. Q. Ai, X. J. Wang, G. T. Liu and L. G. Liu, Phys. Rev. E,
68, 061105 (2003); B. Q. Ai, X. J. Wang, G. T. Liu and L. G. Liu,
Phys. Rev. E, 67, 022903 (2003); B. Q. Ai, G. T. Liu, H. Z. Xie,
L.G. Liu, Chaos 14(4),957 (2004).
\bibitem{8}M. O. Magnasco, Phys. Rev. Lett. 71,  1477 (1993).
\bibitem{9}P. Hanggi and R. Bartussek, Nonlinear physics of
complex system - Current status and Future Trends, 476, Spring,
Berlin, (1996), 294.
\bibitem{10}P. Reimann, R. Bartussek, R. Haussler and P. Hanggi,
Phys. Lett. A  215,  26 (1994).
\bibitem{11}C. R. Doering, W. Horsthemke and J. Riordan, Phys. Rev.
Lett. 72,  2984 (1994).
\bibitem{12}D. Reguera, and J. M. Rubi, Phys. Rev. E 64, 061106
(2001).
\bibitem{13}D. Reguera, G. Schmid, P. S. Burada, J. M. Rubi,
P. Reimann, and P. Hanggi, Phys. Rev. Lett 96, 130603 (2006).
\bibitem{14}M. Badoual, F. Julicher and J. Prost, PNAS 99, 6696
(2002)

\bibitem{15}R. Zwanzig, J. Phys. Chem. 96, 3926 (1992).
\bibitem{16}M. H. Jacobs, Diffusion Processes (Springer, New York,
1967).
\bibitem{17}J. D. Huang, S. T. Brady, B. W. Richards, D. Stenolen,
J. H. Resau, N. G. Copeland and N. A. Jenkins, Nature  397, 204
(1999).
\bibitem{18}P. Gates, K. Cooper, J. Rae, and R. Eisenberg, Prog.
Biophys. Mol. Biol. 53, 153 (1990).
\bibitem{19}M. Fixman, J. Chem. Phys. 69, 1527 (1978).
\bibitem{20}R. A. Seigel, J. Controlled Release 69, 109 (2000).
\bibitem{21}J. P. K. Doye and D. Frenkel, J. Chem. Phys. 110, 7073
(1999).


\end{thebibliography}
\end{document}